\documentclass[prb,aps,showpacs]{revtex4}

\begin{document}

\title{On the origin of unusual transport properties observed in densely packed polycrystalline $CaAl_{2}$}
\author{S. Sergeenkov}
\affiliation{Centro Multidisciplinar para o Desenvolvimento de
Materiais Ceramicos, Departamento de F\'{i}sica, Universidade
Federal de
S\~{a}o Carlos, 13565-905 S\~{a}o Carlos, SP, Brazil and\\
Joint Institute for Nuclear Research, Bogoliubov Laboratory of
Theoretical Physics, 141980 Dubna, Moscow region, Russia}

\author{F.M. Araujo-Moreira}
\affiliation{Centro Multidisciplinar para o Desenvolvimento de
Materiais Ceramicos, Departamento de F\'{i}sica, Universidade
Federal de S\~{a}o Carlos, 13565-905 S\~{a}o Carlos, SP, Brazil}

\date{\today}

\begin{abstract}
A possible origin of unusual temperature behavior of transport
coefficients observed in densely packed polycrystalline $CaAl_{2}$
compound [M. Ausloos et al., J. Appl. Phys. {\bf 96}, 7338 (2004)]
is discussed, including a power-like dependence of resistivity
with $\rho \propto T^{-3/4}$ and $N$-like form of the thermopower.
All these features are found to be in good agreement with the
Shklovskii-Efros localization scenario assuming polaron-mediated
hopping processes controlled by the Debye energy.
\end{abstract}


\maketitle

After the discovery of superconductivity in $MgB_2$, a search
began for other intermetallic compounds with similar structure or
lattice symmetry  (see, e.g.~\cite{1,2,3,4,5,6} and further
references therein). In particular, a class of pseudoternary
compounds $CaAl_{2-x}Si_x$ with $C32$ structure is
shown~\cite{3,5} to exhibit superconducting behavior for $x>0.5$.
The structural and thermodynamic investigations confirmed a
BCS-type pairing mechanism in these compounds with a rather strong
electron-phonon coupling~\cite{6}. At the same time, using an
original route (which allowed formation of crystalline rather than
glassy phase), tiny crystals of $CaAl_{2}$ compound (a close
partner of $MgB_2$ but with $MgCu_{2}$-type $C15$ structure) have
been obtained~\cite{7}  ({\it directly} from $Ca_{2}Al_{3}$ phase
which has an eutectic point at $550^{\circ}C$) and packed into a
granular material (with single-phase granules ranging between $10$
and $50\mu m$). Though no tangible superconducting signals were
detected, a rather unusual transport properties have been found in
these interesting compounds (see Ref.~\cite{7} for more details on
sample preparation and actual measurements). In particular, the
electrical resistance clearly exhibits a power-like (rather than
exponential, expected for conventional low-temperature
localization scenarios~\cite{8,9}) temperature behavior with
resistivity $\rho \propto T^{-3/4}$ for $15K<T<70K$ (Fig.3
in~\cite{7}), and decreasing almost linearly between $70$ and
$235K$ (Fig.2 in~\cite{7}). While the measured thermoelectric
power (TEP) $Q(T)$ has a well-defined $N$-like form (Fig.4
in~\cite{7}) with $Q(T)\propto T^{1/2}$ below $60K$, $Q(T)\propto
T^{3/4}$ above $100K$, and $Q(T)$ almost linearly {\it decreasing}
with $T$ in the intermediary regime (Fig.5 in~\cite{7}).

Turning to the interpretation of the above experimental results,
we notice that the very fact that the resistance data do not
follow conventional localization scenarios dominated by a
variable-range-hopping (VRH) mechanism with resistivity
$\rho(T)=\rho_0 \exp[(T_0/T)^p]$ (leading to either
Mott-like~\cite{8} or Shklovskii-Efros-like~\cite{9} behavior with
$p=1/4$ and $1/2$, respectively) apparently hints at a relatively
small value of the characteristic temperature $T_0$ in this
material (so that $T_0/T\ll 1$ for the whole temperature interval
$15K<T<300K$). This, in turn, implies the importance of
electron-phonon interaction effects on the hopping processes when
localization is accompanied by formation of polarons (strongly
polarized regions around electrons in the conduction
band~\cite{10}). Recall that the binding energy of the polaron a
distance $R$ apart from a donor (or acceptor) site is given by
$E=\alpha /4R=\hbar ^2/2m_pa^2$ where $\alpha =e^2/4\pi \epsilon
_0\epsilon $ with $\epsilon$ being the static dielectric
permeability of the polarized crystal, $m_p$ is an effective
polaron mass, and $a=\hbar ^2/m\alpha$ is the polaron size. At low
temperatures (when the principal processes are dominated by the
Debye energy $k_B\theta _D$), $E\simeq k_B\theta _D$ leading to
scattering of phonons with "heavy" polarons (typically~\cite{8},
$m_p\simeq 10m_e$ where $m_e$ is a free carrier mass) implying a
huge value of the dielectric permeability (for example, in doped
titanates $\epsilon \simeq 1000$ leading to $a\simeq 30\AA$). If
we accept this argument, we will have to assume that in addition
to the conventional thermally activated hopping between the
neighboring unoccupied sites governed by correlated VRH-like
processes with conductivity $\sigma _h(T,E)=\sigma
_{0}(E/k_BT)e^{-U}$ (where $U=2R/a+E/k_BT$ with $R$ being the
hopping distance, $a$ localization length, and $E=\alpha /4R$ an
energy difference between two localized states; $\sigma _0=4\nu
e^2/\alpha $ with $\nu$ being a characteristic phonon frequency),
we are dealing with the so-called phonon-assisted mechanism of
metal-insulator transition (known to be active in slightly doped
semiconductors with impurity conduction and other disordered
systems~\cite{8,9}) which is a hopping process substantially
modified by electron-phonon interaction controlled by the Debye
temperature $\theta _D$. (It is worth mentioning a somewhat
similar mechanism in slightly doped manganites where spin polaron
hopping is controlled by the exchange energy~\cite{11,12}.) At
high temperatures (for $T>\theta _D/2$), this contribution to the
observed conductivity has a thermally activated form of $\sigma
_{th-ph}(T,E)=\sqrt{\frac{\theta
_D}{2T}}\sqrt{\frac{E}{2E_a}}\sigma _h(T,E)$ where $E_a\equiv
E(R=a)=\alpha /4a$, while at low temperatures (for $T<\theta
_D/2$) the conductivity is governed by the phonon-assisted polaron
hopping with $\sigma _{d-ph}(E)=\sigma
_{th-ph}(T=\frac{1}{2}\theta _D,E)$. As we shall see below, the
latter contribution dominates the temperature behavior of the
resistivity and TEP data under discussion. Let us start with the
resistivity. According to the above-mentioned scenario the
observed temperature dependence of $\rho$ should follow the law:
\begin{equation}
\rho (T)=\left [\sigma ^{-1}(T,E)\right ]_{E=E_0(T)}
\end{equation}
where $\sigma (T,E)=\sigma _h(T,E)+\sigma _{th-ph}(T,E)+\sigma
_{d-ph}(E)$, and $E_0(T)$ is defined via the temperature
dependence of the minimal hopping distance $R_0(T)$. The latter is
the solution of the extremum equation $dU(R)/dR=0$ where
$U(R)=\frac{2R}{a}+\frac{\alpha}{4k_BTR}$. Hence,
$R_0(T)=(a/4)\sqrt{T_0/T}$, $E_0(T)\equiv
E(R_0)=(k_BT_0/2)\sqrt{T/T_0}$, $U_0(T)\equiv
U(R_0)=\sqrt{T_0/T}$, and $k_BT_0=2\alpha /a$. As a result, Eq.(1)
can be written as follows
\begin{equation}
\rho (T)=\rho _0\left (\frac{T}{T_0}\right )^{-3/4}\left [1+\delta
_1\left (\frac{T}{T_0}\right )^{5/4}e^{\sqrt{T_0/T}}+\delta
_2\left (\frac{T}{T_0}\right )^{3/2}e^{\sqrt{T_0/T}}\right ]
\end{equation}
Here $\rho _0=8(\theta _D/T_0)e^{\gamma}\sigma _0^{-1}$, $\delta
_1=(T_0/4\theta _D)e^{-\gamma}$, and $\delta _2=(2T_0/\theta
_D)^{3/2}e^{-\gamma}$ with $\gamma =\sqrt{2T_0/\theta _D}$. As we
shall see below, for $15K<T<70K$ and the estimates of the model
parameters ($T_0$, $\theta _D$, and $\delta _{1,2}$), the second
and third terms in the rhs of Eq.(2) can indeed be regarded as
small corrections to the main $T^{-3/4}$ dependence found to
dominate the observed resistivity.

Turning to the TEP data, we notice that like the previously discussed
resistivity, the temperature dependence of the thermoelectric power
$Q(T)$ will be controlled by the hopping energy $E_0(T)$ as well, so
that
\begin{equation}
Q(T)=3Q_c(T)+\frac{\pi ^2k_B^2T}{3e}\left [\frac{d\ln \sigma
(T,E)}{dE}\right ]_{E=E_0(T)}
\end{equation}
Here the first term accounts for concentration dependent
contribution of the three processes with $Q_c(T)=Q_0\ln(N_v/n)$
where $Q_0=\pi ^2k_B/3e$, $N_v(T)=(2\pi m_pk_BT/\hbar ^2)^{3/2}$,
$N$ is the number of sites, and $n\simeq \sqrt{NN_v}$ is the
carriers (polarons) number density. Using the above-introduced
definition of $\sigma (T,E)$, for the temperature range under
discussion, from Eq.(3) we obtain
\begin{equation}
Q(T)=3Q_c(T)+2Q_0\left[\sqrt{\frac{T}{T_0}}+A\left
(\frac{T}{T_0}\right)^{-3/4}+B\left(\frac{T}{T_0}\right
)^{3/4}\right]
\end{equation}
for the temperature behavior of the observed TEP. Here $A=\theta
_D/4T_0$ and $B=\sqrt{T_0/\theta _D}$. It can be easily verified
that, in agreement with the observations, the above expression for
$Q(T)$ exhibits a maximum at $T_{max}=(3A/2)^{4/5}T_0$ and a
minimum at $T_{min}=(A/B)^{2/3}T_0$. In view of the definition of
the coefficients $A$ and $B$ and using the experimental value of
the minimum temperature ($T_{min}=100K$), we immediately obtain a
reasonable estimate for the Debye temperature in this material,
namely $\theta _D=4^{2/3}T_{min}\simeq 250K$ (recall that for
$CaAlSi$ $\theta _D\simeq 226K$~\cite{5,6}). Furthermore, using
the observed values of the maximum temperature $T_{max}=60K$ and
the corresponding TEP extrema, $Q_{max}\equiv Q(T_{max})=23\mu
V/K$ and $Q_{min}\equiv Q(T_{min})=17\mu V/K$, as well as an
obvious relation between the concentration contributions,
$Q_c(T_{min})=Q_c(T_{max})+(3/4)Q_0\ln(T_{min}/T_{max})$, we
obtain the following estimates of the model parameters: $T_0=10K$,
$A=6.25$, $B=0.2$, $\gamma =0.28$, $\delta _1=0.0075$, $\delta
_2=0.015$, $c_{max}=N_v(T_{max})/N=0.04$, and
$c_{min}=N_v(T_{min})/N=0.02$. It is worth mentioning that the
latter estimates of concentrations (with $c\ll 1$) provide further
evidence in favor of adopted here polaron concept~\cite{8} for
explanation of the observed $N$-like TEP form. Returning to the
resistivity, let us notice that the above estimates corroborate
our conjecture about dominant character of the observed $\rho(T)
\propto T^{-3/4}$ law for $15K<T<70K$ temperature interval
providing at the same time an estimate of the characteristic model
conductivity $\sigma _0=16\pi \epsilon \epsilon _0\nu =2\times
10^6\Omega ^{-1}cm^{-1}$ which gives $\nu =4\times 10^{14}s^{-1}$
for phonon frequency (using $\epsilon =1000$). Besides, at high
temperatures the model predicts a small increase of resistivity as
$\rho(T)\simeq \rho_0\delta _1\sqrt{T/T_0}$, in agreement with the
observations. Finally, the deduced estimate of the localization
temperature $T_0=2\alpha a/k_B=10K$ gives $a \simeq 40 \AA$,
$m_p=(\hbar /\alpha )^2(k_BT_0/2) \simeq 8m_e$ and $R_0=a/4 \simeq
10 \AA$ for the polaron size, polaron mass and hopping distance,
respectively, and explains why the Shklovskii-Efros law $\rho
(T)\propto \exp(\sqrt{T_0/T})$ is not seen in the resistivity data
for the whole temperature interval $15K<T<300K$.

In conclusion, a brief comment is in order on the role of
grain-boundary effects in the transport anomalies under
discussion. Polarized light microscopy analysis of this densely
packed polycrystalline material revealed (Fig.1 in~\cite{7}) a
dendritic structure with a well-defined crystalline phases within
a single grain. Besides, the very fact that the adopted here
polaron picture reasonably well describes {\it both} electric
resistivity and thermoelectric power suggests a rather high
quality of this material (which is also evident from its X-ray
diagram shown in Fig.7 from Ref.~\cite{7}) with presumably narrow
enough grain distribution and quasi-homogeneous low-energy
barriers between the adjacent grains.

We gratefully acknowledge financial support from Brazilian agency
FAPESP.

\end{document}